\documentclass[a4paper,12pt]{spieman}  % use this instead for A4 paper
\usepackage{amsmath,amsfonts,amssymb}
\usepackage{caption}
\usepackage{graphicx}
\usepackage{booktabs}
\usepackage{setspace}
\usepackage{tocloft}
\usepackage{lipsum}
\usepackage{siunitx}
\usepackage{tcolorbox}
\usepackage[export]{adjustbox}
\usepackage{lineno}
\title{First Experimental Evaluation of a High-Resolution Deep Silicon Photon-Counting Sensor}

\author[a,b,*]{Rickard Brunskog}
\author[a,b]{Mats Persson}
\author[a]{Zihui Jin}
\author[a,b,**]{Mats Danielsson}
\affil[a]{The Royal Institute of Technology Stockholm, Physics of Medical Imaging, Roslagstullsbacken 21, Stockholm, Sweden, 10691}
\affil[b]{MedTechLabs, BioClinicum, Karolinska University Hospital, Solna, Sweden, 17164}

\cftpagenumbersoff{figure}
\cftpagenumbersoff{table} 
\begin{document} 
\maketitle

\begin{abstract}\\
\textbf{Purpose:} 
Current photon-counting computed tomography detectors are limited to a pixel size of around \SI{0.3}{mm}-\SI{0.5}{mm} due to excessive charge sharing degrading the dose efficiency and energy resolution as the pixels become smaller. In this work, we present measurements of a prototype photon-counting detector that leverages the charge sharing to reach a theoretical sub-pixel resolution in the order of \SI{1}{\micro\metre}. The goal of the study is to validate our Monte-Carlo simulation using measurements, enabling further development.\\
\textbf{Approach:} We measure the channel response at the MAX~IV Lab, in the DanMAX beamline, with a \SI{35}{keV} photon beam, and compare the measurements with a 2D Monte Carlo simulation combined with a charge transport model. Only a few channels on the prototype are connected to keep the number of wire bonds low.\\
\textbf{Results:} The measurements agree generally well with the simulations with the beam close to the electrodes but diverge as the beam is moved further away. The induced charge cloud signals also seem to increase linearly as the beam is moved away from the electrodes.\\
\textbf{Conclusions:} The agreement between measurements and simulations indicates that the Monte-Carlo simulation can accurately model the channel response of the detector with the photon interactions close to the electrodes, which indicates that the unconnected electrodes introduce unwanted effects that need to be further explored. With the same Monte-Carlo simulation previously indicating a resolution of around \SI{1}{\micro\metre} with similar geometry, the results are promising that an ultra-high resolution detector is not far in the future.

\end{abstract}

% Include a list of up to six keywords after the abstract

% Include email contact information for the corresponding author
{\noindent \footnotesize\textbf{*}Rickard Brunskog,  \linkable{rbru@kth.se}\\ \footnotesize\textbf{**}Mats Danielsson,  \linkable{md@mi.physics.kth.se}}\\
%{\noindent \footnotesize\textbf{*}Mats Danielsson, ~ \linkable{md@mi.physics.kth.se}}

%\begin{spacing}{2}   % use double spacing for rest of manuscript

\section{Introduction}
\label{sect:intro}  % \label{} allows reference to this section

Since the first commercial computed tomography scanner was introduced in 1972, many innovations have been made, one of the more recent being the photon-counting detector. A photon-counting detector uses direct conversion from the photon to the generated electric pulse by measuring the current induced on the electrodes by the electron-hole pairs created by the photon interaction. This enables the counting of each photon and labelling the interaction with its corresponding energy, resulting in higher resolution, mitigation of beam hardening artefacts, potentially lowering the dose to the subject, and material basis decomposition with a single X-ray source\cite{PCCTReview,review1,dualCT,highResolution,optimalWeighting,pcd1,pcd2,pcd3,pcd4,pcd5,pcd6}.\\
An integral part of a clinically viable detector is the spatial resolution, which relies heavily on the pixel size. In conventional detectors, having a pixel size in the range of \SI{300}{\micro\metre} to \SI{1000}{\micro\metre}, charge sharing occurs if the interaction happens at the pixel boundary and results in potential double counting of the X-rays unless properly handled by anti-coincidence logic. However, the charge sharing can be leveraged to achieve sub-pixel resolution by estimating the exact interaction position of the photon by observing collected charges on adjacent pixels. This technique, which is similar to the technique used in Anger cameras, has been demonstrated in the MÖNCH detector to reach a resolution close to the micron level. The MÖNCH detector is a face-on hybrid detector and is designed for energies $E\lesssim \SI{20}{keV}$ and count rates of around $10^5\SI{}{s^{-1}}$, both too low for medical diagnostic imaging which falls in the range of \SI{20}{keV}-\SI{150}{keV} and $10^7\SI{}{s^{-1}}$ respectively.\cite{nuclear,mönch}\\ 
To reach an adequate dose efficiency for these energies using silicon, an edge-on design is preferred. In the edge-on orientation, the silicon is oriented such that the plane of the silicon is parallel to the X-rays, having the electrodes placed in successive rows, with the wafer thickness giving the resolution in the orthogonal direction\cite{siliconPCD}.\\
We are developing an edge-on, deep-silicon detector, with a pixel size of $14\times650\SI{}{\micro\metre}$ ($width\times height$), capable of leveraging the charge sharing to achieve sub-pixel resolution in the order of \SI{1}{\micro\metre}. The pixel asymmetry stems from the chosen sensor material, silicon, which needs to be in an edge-on orientation to achieve acceptable dose efficiency. Further, note that the resolution of \SI{1}{\micro\metre} is only in one dimension. The detector can not reach the same resolution in the orthogonal dimension but simulation results indicate around \SI{80}{\micro\metre} using location estimation algorithms, see Sundberg et~al.~(2021). With no estimation algorithm, the orthogonal resolution is given by the sensor thickness. In this prototype, each depth segment in Fig.~\ref{fig:prototypeAndSensor}(a) is \SI{500}{\micro\metre}, which means that for the connected pixels the sensitive volume is ideally $14\times650\times500\SI{}{\micro\metre\cubed}$ ($width\times height\times depth$). The staggered pattern is to fit the connection pad between the pixel strips and the wire bonds, the electrode strips themselves are parallel to the incoming X-rays with each depth segment containing 384 adjacent electrodes.\cite{oneMicron}\\
In this work, we present experimental data of the first prototype of the aforementioned detector, taken at the DanMAX beamline at the MAX IV Laboratory in Lund, Sweden. Further, we compare the experimental data with simulated data by combining Monte Carlo photon simulation with a charge transport model.\cite{penelope}

\section{Method}
\begin{figure}[t]
     \centering
            \includegraphics[width = 0.98\textwidth]{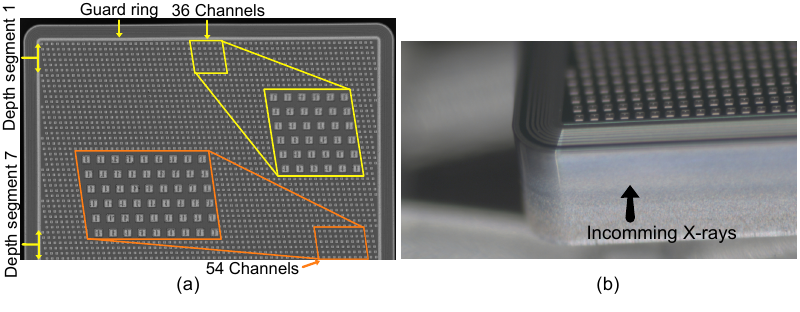}
        \caption{(a) Photograph of the prototype sensor from above, with the 36 bonded channels in the first depth segment marked, as well as the 54 bonded channels in the 7th depth segment. (b) Photograph of the sensor front indicating the photon direction.}
        \label{fig:prototypeAndSensor}
\end{figure}

We are developing a novel, edge-on, single-sided, segmented silicon strip detector which, from previous work done by Sundberg et~al.~(2021), has shown a theoretical resolution of around \SI{1}{\micro\metre} using a similar geometry. An image of the prototype sensor, fabricated by GE HealthCare, from above and an image of the front are shown in Fig.~\ref{fig:prototypeAndSensor}(a) and Fig.~\ref{fig:prototypeAndSensor}(b) respectively.
Since the purpose of this study is to validate the Monte-Carlo simulation, and thus in extension to evaluate the viability of this detector, we used an existing ASIC, see the work of Gustavsson et~al. (2012), and Xu~et~al. (2013). While not optimised for the purpose of this study, it offers a chance to obtain early data as input to a future ASIC design optimised for the high-resolution sensor. Each ASIC has 160 channels, and 8 thresholds configurable to DAC levels from 0 to 255. The prototype sensor has 36 wire-bonded channels in the middle of the first depth segment and 54 wire-bonded channels at the edge of the seventh depth segment, marked in Fig.~\ref{fig:prototypeAndSensor}(a), with the channel electrodes set to \SI{0}{\volt} and the backside to \SI{250}{\volt}. The measurements are performed with the 36 bonded channels in the first depth segment. Before each measurement, the channels are calibrated such that \SI{0}{keV} corresponds to DAC-level 20.\cite{oneMicron,asic0, asic1}
\subsection*{Experiment}
\begin{figure}[t]
     \centering
            \includegraphics[width = 0.98\textwidth,valign=c]{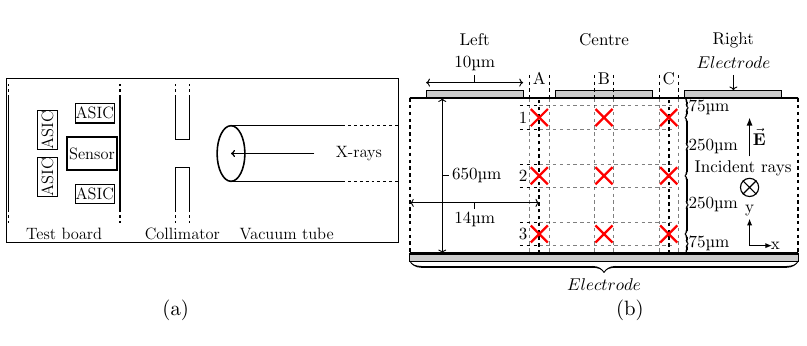}
        \caption{(a) Outline of the experiment setup. (b) Front view of three pixels on the sensor with beam positions $A_1$ through $C_3$ with the X-rays going into the plane of the figure.}
        \label{fig:sensorAndSetup}
\end{figure}
The experiment is performed in the DanMAX beamline at the MAX IV Laboratory in Lund, Sweden. The beam is collimated by \SI{2}{\milli\metre} thick tungsten slits configured to a $10\times10$~\SI{}{\micro\metre} opening placed in between the vacuum tube and the sensor, with the beam profile measured by an X-ray beam viewer\footnote{\label{beamviewer}The beam viewer is provided by the DanMAX beamline and is based on a YAG scintillating screen, with a reversed Navitar 35-mm camera lens and a Basler GigE acA1300-30gm camera, with an effective pixel size of \SI{2.4}{\micro\metre}.} mounted in the beam path, which is then removed before the measurements.\\
The testing board with the sensor is mounted on top of a hexapod with a resolution of $\pm\SI{0.5}{\micro\metre}$ and a repeatability of $\pm\SI{0.5}{\micro\metre}$. The experimental setup is shown in Fig.~\ref{fig:sensorAndSetup}(a).\\
To align the sensor with the beam, the sensor is first visually positioned using the X-ray beam viewer mounted behind the prototype, after which the channel response is observed as the sensor is scanned across the beam to find the position with the maximum number of counts in the chosen channel, the beam should then be at the centre of the channel in the scanned dimension. The sensor is aligned in the directions perpendicular to the beam, as well as rotation-wise around the axes.\\
The sensor is positioned with the beam in the marked positions (red crosses) in Fig.~\ref{fig:sensorAndSetup}(b), and a DAC-sweep is performed at each location, with a frame time of \SI{0.5}{s} per step, the incoming photons at \SI{35}{\kilo e\volt}, and a pulse detection time of \SI{120}{ns}. The energy scale is then calibrated by observing the DAC-sweep in the positions $A_1$, $B_1$, and $C_1$ for the left, centre, and right channels respectively, according to 
\begin{equation}\label{eq:energyCal}
    \dfrac{\Delta_E}{\Delta_{DAC}} = \dfrac{\SI{35}{keV}}{(\mathcal{N}_N-20)~\SI{}{DAC}},
\end{equation}
where $\mathcal{N}_N$ is the DAC-level where the number of counts in the DAC-sweep drops below a threshold $N$ where \SI{35}{keV} is (later) determined to be for the measurements just mentioned.\\ With the beam so close to the electrodes, it is expected that some interactions will be mainly detected by the closest electrodes, and we should see the number of counts drop sharply before flattening out from pile-up, and \SI{35}{keV} should be at, or close to, the DAC-level just before the DAC-sweep flattening out. \\
To ascertain that pile-up would not pose a problem, the beam is positioned in $B_1$ and attenuated using a \SI{75}{\micro\metre} tungsten film. The maximum number of counts above the noise level (around 6\SI{}{keV}), using a frame time of \SI{0.1}{s}, is approximately \SI{6000}{cps}. Removing the attenuation we should therefore see a maximum incoming count rate of around \SI{53000}{cps} at \SI{35}{keV}. This is well below the count rate capabilities of the used ASICs, thus the pile-up can be neglected at most energies below \SI{35}{keV} and we should see the effect only closer to \SI{35}{keV}.\cite{asic0,asic1}\\
The lost charge cloud signal, $\Delta \mathcal{S}$, is then defined as the energy lost to other electrodes as the beam is moved down in the columns of Fig.~\ref{fig:sensorAndSetup}, and is given according to 
\begin{equation}\label{eq:csDef}
    \Delta \mathcal{S}(X_y) = \dfrac{E(X_1)_{N}-E(X_y)_{N}}{E(X_1)_{N}},
\end{equation}
where $E(X_y)_N$ is the energy at which the DAC-sweep counts drop below $N$ as defined in Eq.~\ref{eq:energyCal}, and $X~=~A,~B,~C$, and $y~=~1,~2,~3$. For simplicity, $N$ is defined by the energy calibrations in Eq.~\ref{eq:energyCal} and is kept constant for the remaining positions.

\subsection*{Simulations}
We simulate the incoming photons using the code system Penelope, with a monochromatic beam of \SI{35}{keV} modelled after the beam profile in Fig.~\ref{fig:beamFits}, with the output from Penelope then used as input in a charge transport model. Due to the prototype only having a few connected channels in the first and seventh depth segments, we use a 2D model in the simulation to observe the channel response. Note here that the charge sharing between depth segments is assumed to be minimal in a fully connected sensor compared to the charge sharing between adjacent pixels in width due to the much larger electrode dimension in depth. Further, the dimension of the electrodes in the depth direction is of the same order as the pixel size of current PCDs and any charge sharing between them can be alleviated by using already established techniques\cite{penelope,oneMicron,chargesharing1,chargesharing2,chargesharing3}.\\
The simulated beam is shifted in the X-direction in Fig.~\ref{fig:sensorAndSetup}(b) until the ratio of the simulated counts between the left and right channels at $E$~\SI{}{keV} is similar to the measured ratio between the left and right channels with the beam in column $B$ after energy calibration of the DAC-levels. The same approach is used for the left and centre channels and the right and centre channels with the beam in columns $A$ and $C$ respectively. This is done at every position to take any positioning and alignment errors into account. The energy level of $E$\SI{}{keV} is somewhat arbitrarily chosen, though it needs to be above the noise level and below the sharp drop in counts.\\
To scale the simulation data to the measurements, the mean-squared error is minimised with respect to the scaling parameter and is given by
\begin{equation}
    \text{MSE}(\lambda) = \dfrac{1}{N}\sum_{k=1}^N(\lambda s_k - m_k)^2,
\end{equation}
where $s_k$, $m_k$, and $\lambda$ are the simulation data point, the measurement data point, and the scaling parameter respectively. The scaling parameter is calculated for each row in column $B$ and is subsequently used for the corresponding rows in columns $A$ and $C$.

\begin{table}[t]
	\begin{minipage}{0.66\linewidth}\vspace{0pt}%
		\centering
		\includegraphics[width = \textwidth]{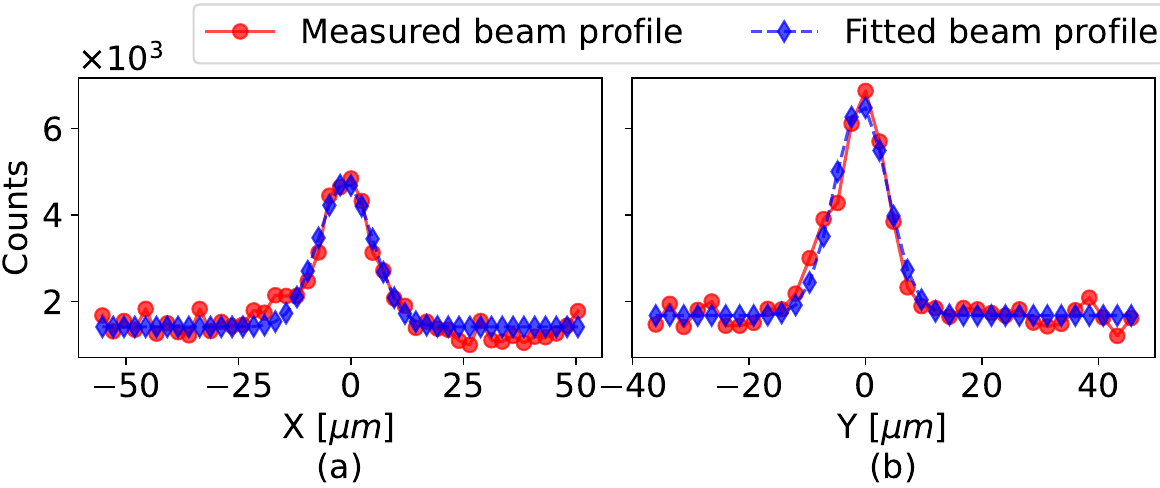}
        \captionof{figure}{Marginal distribution of the beam profile measured with the beam viewer with fitted profiles of the simulated beam.\\(a) $\sigma_x = \SI{6.06}{\micro\metre}$, (b) $\sigma_y = \SI{4.57}{\micro\metre}$.}
        \label{fig:beamFits}
	\end{minipage}
    \begin{minipage}{0.33\linewidth}\vspace{-2.11cm}%
		\caption{Simulation offsets fitted to match measurements}
		\label{tab:offsets}
		\centering
		\begin{tabular}{|c|c|c|c|}
        \hline
         [\SI{}{\micro\metre}] & A & B & C \\
         \hline
        1 & -1.3 & -1.20 & -0.7 \\
        \hline
        2 & -1.6 & -1.75 & -1.3 \\
        \hline
        3 & -1.3 & -0.55 & -0.4 \\
        \hline
    \end{tabular}
	\end{minipage}\hfill
\end{table}
\begin{figure}[ht]
    \centering
    \includegraphics[width = \textwidth]{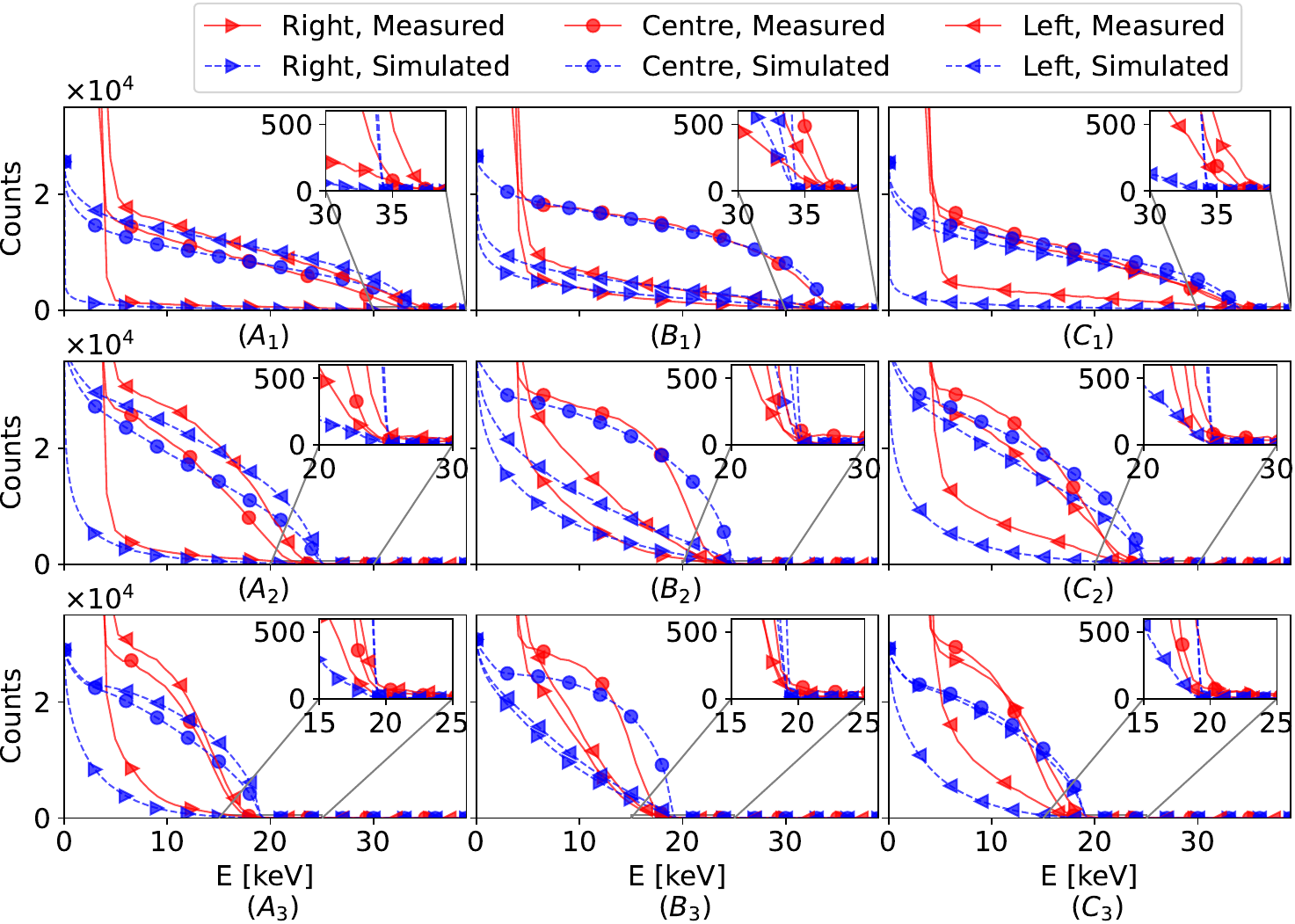}
    \caption{Comparison of the channel responses between measurements and simulations. From the top left in order is each position marked in Fig.~\ref{fig:sensorAndSetup}(b). The channels used are three of the connected channels in the first depth segment in Fig.~\ref{fig:prototypeAndSensor}(a).}
    \label{fig:fullComparison}
\end{figure}

\section{Results}
The measured beam profile is shown in Fig.~\ref{fig:beamFits}, along with the fitted profile with $\sigma_x=\SI{6.06}{\micro\metre}$ and $\sigma_y=\SI{4.57}{\micro\metre}$ used in the simulations.\\
Table~\ref{tab:offsets} shows the resulting offsets used in the simulations to take positioning uncertainties into account with the energy $E$~\SI{}{keV} chosen by visual inspection of the graphs in Fig. \ref{fig:fullComparison} to be \SI{10}{keV}.\\
The DAC-level where the number of photon counts drops below $N~=~500$ is chosen as the reference point for \SI{35}{keV} in the energy calibration of Eq.~\ref{eq:energyCal} and the lost charge cloud signal of Eq.~\ref{eq:csDef}. This point is after a sharp drop in the number of counts, and slightly after the curve starts to flatten out due to the pile-up, as can be seen in the zoomed-in windows of the first row of graphs in Fig.~\ref{fig:fullComparison}. The energy calibration then yields \SI{0.825}{keV/DAC}, \SI{0.814}{keV/DAC}, and \SI{0.747}{keV/DAC} for the right, centre, and left channels respectively.\\
Figure \ref{fig:fullComparison} shows the results from the DAC-sweeps in each marked position of Fig.~\ref{fig:sensorAndSetup}(b) as well as zoomed-in windows on the calibration point.\\ 
The scaling parameter is calculated using the data points between \SI{7}{keV} and \SI{30}{keV} for row 1, \SI{7}{keV} and \SI{21}{keV} for row 2, and \SI{7}{keV} and \SI{17}{keV} for row 3. The decrease of the upper calibration point is due to charge-sharing effects which move the calibration point down on the energy scale as the beam moves from row 1 to row 3, as can be seen in Fig.~\ref{fig:fullComparison}.\\
Table \ref{tab:drop} shows the number of counts at $E\approx \SI{10}{keV}$ for the centre channel with the beam in positions $B_1$, $B_2$, and $B_3$.\\ Figure~\ref{fig:LCCS} shows the resulting $\Delta \mathcal{S}$ as defined by Eq.~\ref{eq:csDef}.

\begin{table}[t]
    \begin{minipage}{0.45\linewidth}\vspace{0cm}%
		\caption{Number of counts at $E~\approx~\SI{10}{keV}$ for measurements and simulations (not scaled) in the centre channel}
		\label{tab:drop}
		\centering
		\begin{tabular}{|c|c|c|c|}
        \hline
          & $B_1$ & $B_2$ & $B_3$ \\
          \hline
        Measurements & $17351$& $26930$& $25869$\\
         \hline
        Simulations & $22855$& $23461$& $25908$ \\
        \hline
    \end{tabular}
	\end{minipage}
	\begin{minipage}{0.5\linewidth}\vspace{0pt}%
		\centering
		\includegraphics[width = \textwidth]{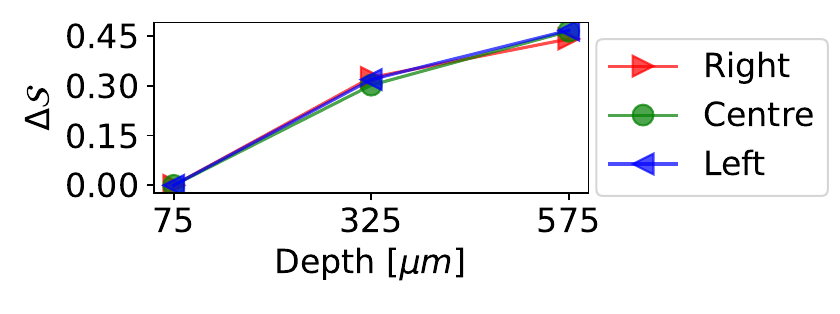}
        \captionof{figure}{$\Delta \mathcal{S}$: Ratio of the lost charge (relative to the positions closest to the electrodes) as a function of beam position, as defined by Eq.~\ref{eq:csDef}.}
        \label{fig:LCCS}
	\end{minipage}
    \hfill
\end{table}

\section{Discussion}

The offsets in Tab. \ref{tab:offsets} are motivated by the uncertainty in the hexapod positioning as well as the alignment. Any other sources of positioning uncertainties are assumed to be accounted for by these offsets or otherwise can be neglected.\\
In Fig.~\ref{fig:fullComparison} one can note that at energies below $E\approx\SI{6}{keV}$ the simulations diverge heavily from the measurements. This is due to the readout electronics not being optimised for the pixel size of the sensor, leading to a high noise floor. This divergence is expected as the Monte-Carlo simulation does not model this noise.\\ As can be seen in Fig.~\ref{fig:fullComparison}, the channel response with the beam in the first row agrees well between simulations and measurements, with only a slight underestimation of the low energy interactions (before the simulations diverge), and a slight overestimation of the high energy interactions. However, when the beam is moved away from the electrodes the agreement falters. It is believed that this stems from the fact that no channels behind the first depth segment are connected, resulting in the unconnected channels having an induced potential, as well as the electric field being distorted from the field in a fully connected sensor.\\
Figure \ref{fig:fullComparison} also shows a drop in the number of counts with the beam in the first row compared to the beam in the second and third rows, with Tab. \ref{tab:drop} showing the number of counts at $E\approx\SI{10}{keV}$ for both measurements and simulations. The measurements show a drop in counts of approximately $35\%$ going from the second and third row to the first row. A similar effect is seen in the simulations, and for the corresponding simulated channels, the drop is approximately $10\%$. The $10\%$ seen in the simulation stems from the charge cloud not spreading out to as many electrodes when the beam is in row 1. The difference between the measurements and the simulations is believed to also stem from the distorted field mentioned above and becomes pronounced as the beam is moved away from the electrodes. The Monte-Carlo simulation is currently not capable of modelling a partially connected sensor with induced potential on unconnected electrodes.\\
Observing the zoomed-in windows of the calibration point shows the simulations and measurements agreeing generally well in all positions, albeit slightly less well with increased $ y$ depth. The resulting $\Delta \mathcal{S}$, shown in Fig.~\ref{fig:LCCS}, has a fairly linear trend, indicating that the estimation of the interaction position will be more accurate if the photon interacts further away, rather than close to the electrodes. It is important to note that $\Delta\mathcal{S}$ in Eq.~\ref{eq:csDef} is almost certainly non-zero in positions $A_1$, $B_1$, $C_1$, but is zero by definition due to having no other point of reference.\\
Single interaction readouts have not yet been performed with the prototype using synchrotron radiation. A prototype, with all unconnected electrodes grounded to provide the field of a fully connected sensor, is being made and we aim to perform these measurements in the near future.\\
As some final remarks, we want to address the field of view and the quantum efficiency. Though the prototype is small, the sensors would be able to be set up in an array. For example, a full computed tomography detector would be comprised of a detector array, where the number of detector modules in the array would be adjusted to cover the required field of view. In the same way, depending on the use case, several detector modules would be able to be stacked behind each other to increase the quantum efficiency.

\section{Conclusion}
We have presented a comparison between simulations and measurements of a prototype deep silicon detector with a pixel size of $14\times650$~\SI{}{\micro\metre}. Whilst the simulation differs increasingly from the measurements as a function of the depth of the beam on the sensor front, the agreement of the channel response with the beam close to the electrodes provides promise that the Monte Carlo simulation can be used to simulate the response of the sensor. Further work includes developing the algorithms for interaction position estimation, and the necessary read-out electronics. Due to the noise being high in the prototype, capturing the full charge cloud becomes difficult as the tails become intermingled with the noise, thus a more rigorous energy calibration can be done once the algorithms and the electronics are in place.\\
Whilst these measurements in and of themselves do not constitute evidence of a resolution of \SI{1}{\micro\metre}, coupled with the agreement with the simulations, they provide confidence that we are on the right track.\\
The potential of a detector with a \SI{1}{\micro\metre} resolution can be used in many applications, such as in synchrotron research and in medical imaging. An intriguing application is the area of phase-contrast imaging, where a high-resolution detector could remove the need for an analyser grating, reducing the complexity of clinical implementation\cite{hardXRay,lowBrilliance,scherer,humanScale,spectralPCI}.

\section{Disclosures}
Rickard Brunskog, Zihui Jin, and Mats Persson disclose research collaboration with GE HealthCare. Mats Persson also discloses research support and license fees from GE HealthCare. Mats Danielsson has an active consulting agreement with GE HealthCare.

\section{Code, Data, and Materials Availability}
The simulation code used in this article is proprietary and is not publicly available.\\
The data used in this article is available upon reasonable request to the authors.

\section{Acknowledgements}
We acknowledge MAX IV Laboratory for time on Beamline DanMAX under Proposal 20220198. Research conducted at MAX IV is supported by the Swedish Research council under contract 2018-07152, the Swedish Governmental Agency for Innovation Systems under contract 2018-04969, and Formas under contract 2019-02496. DanMAX is funded by the NUFI grant no. 4059-00009B.\\
We would also like to thank Innokenty Kantor, senior researcher at the DanMAX beamline, for helping make sure the experiment went smoothly.\\
This study received financial support from the Olle Engkvist Foundation grant 202-0702 and from MedTechLabs.\\
The data for the centre channel has been used in an SPIE conference abstract. The abstract is currently under consideration and has the title "\textit{Experimental Evaluation of a Micron-Resolution CT Detector}". 

%%%%% References %%%%%
\bibliography{report}   % bibliography data in report.bib

\begin{thebibliography}{10}

\bibitem{PCCTReview}
T.~Flohr, M.~Petersilka, A.~Henning, {\em et~al.}, ``Photon-counting ct review,'' {\em Physica Medica: European Journal of Medical Physics} {\bf 79}, 126--136  (2020).
\newblock doi: 10.1016/j.ejmp.2020.10.030.

\bibitem{review1}
M.~Danielsson, M.~Persson, and M.~Sjölin, ``Photon-counting x-ray detectors for ct,'' {\em Physics in Medicine \& Biology} {\bf 66}, 03TR01  (2021).

\bibitem{dualCT}
S.~Faby, S.~Kuchenbecker, S.~Sawall, {\em et~al.}, ``Performance of today's dual energy ct and future multi energy ct in virtual non-contrast imaging and in iodine quantification: A simulation study,'' {\em Medical Physics} {\bf 42}(7), 4349--4366  (2015).

\bibitem{highResolution}
M.~Mannil, T.~Hickethier, J.~von Spiczak, {\em et~al.}, ``Photon-counting ct: High-resolution imaging of coronary stents,'' {\em Investigative Radiology} {\bf 53}(3), 143--149  (2018).

\bibitem{optimalWeighting}
T.~G. Schmidt, ``Optimal “image-based” weighting for energy-resolved ct,'' {\em Medical Physics} {\bf 36}(7), 3018--3027  (2009).

\bibitem{pcd1}
A.~M. Alessio and L.~R. MacDonald, ``Quantitative material characterization from multi-energy photon counting ct,'' {\em Medical Physics} {\bf 40}(3), 031108  (2013).

\bibitem{pcd2}
R.~J. Acciavatti and A.~D.~A. Maidment, ``A comparative analysis of otf, nps, and dqe in energy integrating and photon counting digital x-ray detectors,'' {\em Medical Physics} {\bf 37}(12), 6480--6495  (2010).

\bibitem{pcd3}
X.~Liu, F.~Grönberg, M.~Sjölin, {\em et~al.}, ``Count rate performance of a silicon-strip detector for photon-counting spectral ct,'' {\em Nuclear Instruments and Methods in Physics Research Section A: Accelerators, Spectrometers, Detectors and Associated Equipment} {\bf 827}, 102--106  (2016).

\bibitem{pcd4}
F.~Grönberg, M.~Danielsson, and M.~Sjölin, ``Count statistics of nonparalyzable photon-counting detectors with nonzero pulse length,'' {\em Medical Physics} {\bf 45}(8), 3800--3811  (2018).

\bibitem{pcd5}
H.~Atak and P.~M. Shikhaliev, ``Photon counting x-ray imaging with k-edge filtered x-rays: A simulation study,'' {\em Medical Physics} {\bf 43}(3), 1385--1400  (2016).

\bibitem{pcd6}
W.~C. Barber, J.~C. Wessel, E.~Nygard, {\em et~al.}, ``High flux energy-resolved photon-counting x-ray imaging arrays with cdte and cdznte for clinical ct,'' in {\em 2013 3rd International Conference on Advancements in Nuclear Instrumentation, Measurement Methods and their Applications (ANIMMA)},  1--5  (2013).

\bibitem{nuclear}
T.~E. Peterson and L.~R. Furenlid, ``{SPECT} detectors: the anger camera and beyond,'' {\em Physics in Medicine \& Biology} {\bf 56}(17), R145  (2011).

\bibitem{mönch}
S.~Cartier, M.~Kagias, A.~Bergamaschi, {\em et~al.}, ``Micrometer-resolution imaging using {MÖNCH}: towards g2-less grating interferometry,'' {\em Journal of Synchrotron Radiation} {\bf 23}(6), 1462--1473  (2016).

\bibitem{siliconPCD}
C.~Xu, H.~Chen, M.~Persson, {\em et~al.}, ``Energy resolution of a segmented silicon strip detector for photon-counting spectral ct,'' {\em Nuclear Instruments and Methods in Physics Research Section A: Accelerators, Spectrometers, Detectors and Associated Equipment} {\bf 715}, 11--17  (2013).

\bibitem{oneMicron}
C.~Sundberg, M.~U. Persson, J.~J. Wikner, {\em et~al.}, ``1-µm spatial resolution in silicon photon-counting ct detectors,'' {\em Journal of Medical Imaging} {\bf 8}(6), 063501  (2021).

\bibitem{penelope}
N.~E. Agency, {\em PENELOPE 2018: A code system for Monte Carlo simulation of electron and photon transport}, OECD iLibrary  (2019).

\bibitem{asic0}
M.~Gustavsson, F.~U. Amin, A.~Bjorklid, {\em et~al.}, ``A high-rate energy-resolving photon-counting asic for spectral computed tomography,'' {\em IEEE Transactions on Nuclear Science} {\bf 59}(1), 30--39  (2012).

\bibitem{asic1}
C.~Xu, M.~Persson, H.~Chen, {\em et~al.}, ``Evaluation of a second-generation ultra-fast energy-resolved asic for photon-counting spectral ct,'' {\em IEEE Transactions on Nuclear Science} {\bf 60}(1), 437--445  (2013).

\bibitem{chargesharing1}
S.~S. Hsieh, ``Coincidence counters for charge sharing compensation in spectroscopic photon counting detectors,'' {\em IEEE Transactions on Medical Imaging} {\bf 39}, 678--687  (2020).

\bibitem{chargesharing2}
X.~Ji, R.~Zhang, G.-H. Chen, {\em et~al.}, ``Impact of anti-charge sharing on the zero-frequency detective quantum efficiency of cdte-based photon counting detector system: cascaded systems analysis and experimental validation,'' {\em Physics in Medicine \& Biology} {\bf 63}, 095003  (2018).

\bibitem{chargesharing3}
H.-E. Nilsson, B.~Norlin, C.~Fröjdh, {\em et~al.}, ``Charge sharing suppression using pixel-to-pixel communication in photon counting x-ray imaging systems,'' {\em Nuclear Instruments and Methods in Physics Research Section A: Accelerators, Spectrometers, Detectors and Associated Equipment} {\bf 576}(1), 243--247  (2007).
\newblock Proceedings of the 8th International Workshop on Radiation Imaging Detectors.

\bibitem{hardXRay}
F.~Pfeiffer, M.~Bech, O.~Bunk, {\em et~al.}, ``Hard-x-ray dark-field imaging using a grating interferometer,'' {\em Nature Materials} {\bf 7}(2), 134--137  (2008).

\bibitem{lowBrilliance}
F.~Pfeiffer, T.~Weitkamp, O.~Bunk, {\em et~al.}, ``Phase retrieval and differential phase-contrast imaging with low-brilliance x-ray sources,'' {\em Nature Physics} {\bf 2}(4), 258--261  (2006).

\bibitem{scherer}
K.~Scherer, {\em Grating-Based X-Ray Phase-Contrast Mammography}, Springer  (2016).

\bibitem{humanScale}
M.~Viermetz, N.~Gustschin, C.~Schmid, {\em et~al.}, ``Dark-field computed tomography reaches the human scale,'' {\em Proceedings of the National Academy of Sciences} {\bf 119}(8), e2118799119  (2022).

\bibitem{spectralPCI}
K.~Mechlem, T.~Sellerer, M.~Viermetz, {\em et~al.}, ``A theoretical framework for comparing noise characteristics of spectral, differential phase-contrast and spectral differential phase-contrast x-ray imaging,'' {\em Physics in Medicine \& Biology} {\bf 65}(6), 065010  (2020).

\end{thebibliography}
\bibliographystyle{spiejour}   % makes bibtex use spiejour.bst

%%%%% Biographies of authors %%%%%

\vspace{2ex}\noindent\textbf{Rickard Brunskog} graduated with an MSc degree in engineering physics from KTH Royal Institute of Technology in 2022, and is currently a PhD student in the Physics of Medical Imaging Division at KTH. His research interests involve the development of ultra-high resolution deep silicon photon counting detectors and their use in clinical phase-contrast imaging.

 \vspace{2ex}\noindent\textbf{Mats Persson} received his PhD in Physics in 2016 from KTH Royal Institute of Technology in Stockholm. After spending three years as a postdoctoral researcher at Stanford University and at General Electric Research Center, he returned to KTH in 2020 where he is now an Assistant Professor of Physics. His research interests are focused on image reconstruction and mathematical performance modelling for photon-counting spectral CT.

 \vspace{2ex}\noindent\textbf{Zihui Jin} 
 Zihui Jin received his MSc degree in Nanotechnology from KTH Royal Institute of Technology in 2023 and is now working in the semiconductor industry. His research interest involves the simulation and evaluation of deep silicon photon-counting CT detectors.
 
 \vspace{2ex}\noindent\textbf{Mats Danielsson} is a professor at KTH Royal Institute of Technology and division leader in Physics of Medical Imaging as well as a program leader at MedTechLabs (Stockholm). He received his PhD in physics for research conducted at CERN and was a postdoc at Lawrence Berkeley National Laboratory. His main research interests are photon counting detectors and X-ray optics and he has 120 publications in JMI, Medical Physics, PMB, Nature, and other respected journals. He also holds more than 100 patents and is a co-founder of Sectra Mamea AB, C-RAD AB, and Prismatic Sensors.
 
% \vspace{1ex}
% \noindent Biographies and photographs of the other authors are not available.

\listoffigures
\listoftables

\end{document}